\documentclass[twocolumn,showpacs,preprintnumbers,amsmath,amssymb,superscriptaddress]{revtex4-1}

\usepackage{graphicx}
\usepackage{graphics}
\usepackage{dcolumn}
\usepackage{bm}

\begin{document}

\title{From local force-flux relationships to internal dissipations and their impact on heat engine performance: The illustrative case of a thermoelectric generator}

\author{Y. Apertet}\email{yann.apertet@gmail.com}
\affiliation{Institut d'Electronique Fondamentale, Universit\'e Paris-Sud, CNRS, UMR 8622, F-91405 Orsay, France}
\author{H. Ouerdane}
\affiliation{Laboratoire CRISMAT, UMR 6508 CNRS, ENSICAEN et Universit\'e de Caen Basse Normandie, 6 Boulevard Mar\'echal Juin, F-14050 Caen, France}
\affiliation{Universit\'e Paris Diderot, Sorbonne Paris Cit\'e, Institut des Energies de Demain (IED), 75205 Paris, France}
\author{C. Goupil}
\affiliation{Laboratoire CRISMAT, UMR 6508 CNRS, ENSICAEN et Universit\'e de Caen Basse Normandie, 6 Boulevard Mar\'echal Juin, F-14050 Caen, France}
\affiliation{Universit\'e Paris Diderot, Sorbonne Paris Cit\'e, Institut des Energies de Demain (IED), 75205 Paris, France}
\author{Ph. Lecoeur}
\affiliation{Institut d'Electronique Fondamentale, Universit\'e Paris-Sud, CNRS, UMR 8622, F-91405 Orsay, France}

\date{\today}

\begin{abstract}
We present an in-depth analysis of the sometimes understated role of the principle of energy conservation in linear irreversible thermodynamics. Our case study is that of a thermoelectric generator (TEG), which is a heat engine of choice in irreversible thermodynamics, owing to the coupling between the electrical and heat fluxes. We show why Onsager's reciprocal relations must be considered locally and how internal dissipative processes emerge from the extension of these relations to a global scale: the linear behavior of a heat engine at the local scale is associated with a dissipation process that must partake in the global energy balance. We discuss the consequences of internal dissipations on the so-called efficiency at maximum power, in the light of our comparative analyses of exoreversibility and endoreversibility on the one hand, and of two classes of heat engines, autonomous and periodically-driven, on the other hand. Finally, basing our analysis on energy conservation, we also discuss recent works which claim the possibility to overcome the traditional boundaries on efficiency imposed by finite-time thermodynamics in thermoelectric systems with broken time-reversal symmetry; this we do by introducing a ``thermal'' thermopower and an ``electrical'' thermopower which permits an analysis of the thermoelectric response of the TEG considering a possible dissymmetry between the electrical/thermal and the thermal/electrical couplings.
\end{abstract}

\pacs{05.70.Ln, 84.60.Rb}
\keywords{}

\maketitle

\section{Introduction}
Conversion of heat into work is an old problem which led to the development of a variety of heat engines. Thermodynamics emerged as an engineer's activity devoted to finding practical solutions to real-life problems but Sadi Carnot's work \cite{Carnot1824} on the optimization of the operation of heat engines initiated the scientific development of thermodynamics. Carnot's main finding boils down to the fact that a heat engine placed between two thermostats at temperatures $T_{\rm hot}$ and $T_{\rm cold}$ respectively ($T_{\rm hot} > T_{\rm cold}$), operates with a heat-to-work conversion efficiency which cannot exceed the so-called Carnot efficiency $\eta_{\rm C} = 1 - T_{\rm cold} / T_{\rm hot}$. The upper bound $\eta_{\rm C}$ may be reached if and only if the thermodynamic cycle followed by the heat engine is reversible, implying that it takes an infinite time to produce work with an efficiency $\eta_{\rm C}$, as reversibility is associated with quasistatic processes.

Efficiency is a measure of the quantity of heat that cannot be made available for work. The relationship between efficiency and irreversibility, though not obvious at the time of Carnot's work \cite{Carnot1824}, came prior to the discovery by Joule \cite{Joule} of the mechanical equivalence of heat, which led to the principle of the conservation of energy. It is interesting to note that though time seems to play no role at all in classical equilibrium thermodynamics, both the first and the second principles of thermodynamics appear to be closely linked to time nonetheless, albeit in a different fashion: conservation of energy follows from invariance with respect to translation in time (a consequence of Noether's first theorem \cite{Noether1,*Noether2}), and irreversibility characterizes a ``preferred direction'', or asymmetry, of time.

The extension and refinement of classical thermodynamics has its roots in engineering problems too: optimization of the production of \emph{power} in nuclear plants~\cite{FFTfathers1,*FFTfathers2,*FFTfathers3} during the 1950s led engineers and scientists to include a time variable in thermodynamics. The finite-time character of actual processes is fundamental in that it embodies energy dissipation hence irreversibility. In the simplest situation, neglecting parasitic heat leaks, dissipation takes place because of the finite thermal conductance of the elements that couple a perfect heat engine to the thermostats. Dissipation negatively impacts on heat-to-work conversion efficiency but it certainly is necessary, hence useful, for power production. The field of finite-time thermodynamics, which took off with the work of Curzon and Ahlborn \cite{Curzon1975}, is about trading efficiency for power: the quantity of interest is no longer the efficiency of heat-to-work conversion but the efficiency at maximum power: ${\eta_{\rm CA} = 1 - \sqrt{T_{\rm cold} / T_{\rm hot}}}$, also named the Curzon and Ahlborn efficiency. For recent reviews on this topic see Refs.~\cite{ReviewFFT1,*ReviewFFT2}.

In a recent publication, Van den Broeck stated that``\emph{the Curzon-Alhborn efficiency is a fundamental result that follows, without approximation, from the theory of linear irreversible thermodynamics}'' \cite{VandenBroeck2005}. This assertion is based on a linear nonequilibrium thermodynamics analysis of three systems: a generic heat engine, a cascade construction, and a tandem construction, which convert the heat that flows through them into work. Assuming a general heat engine operating between two thermal reservoirs, at temperatures $T_{\rm hot}$ and $T_{\rm cold}$ respectively (in our notation), Van den Broeck described the heat-to-work conversion process with a linear force-flux formalism \cite{Onsager1931}:

\begin{equation}\label{Onsager}
\left(
\begin{array}{c}
J_1\\
J_2\\
\end{array}
\right)
=
\left(
\begin{array}{cc}
L_{11} ~ & L_{12} \\
L_{21}~ & ~L_{22} \\
\end{array}
\right)
\left(
\begin{array}{c}
X_1\\
X_2\\
\end{array}
\right),
\end{equation}

\noindent where $J_1$ is the time derivative of an extensive variable $x$, which is the thermodynamic conjugate of a force $F$ applied to the engine, and $J_2$ is the heat flux. The quantities $X_1$ and $X_2$ are the thermodynamic forces \cite{Callen1985}. The matrix elements $L_{ij}$ are the kinetic coefficients. Now, a question naturally comes to mind: as the thermal contacts between the heat engine and the temperature reservoirs are assumed to be perfect, what is the process that ensures causality for this system? The first principle of thermodynamics, sometimes overlooked in thermodynamic analyses, provides a clear answer to this question. To illustrate our analysis we consider a thermoelectric generator as the heat engine: this class of engines provides a sound physical picture of the coupled processes at work hence permitting a rigorous and fine study of irreversible thermodynamics \cite{deGroot}. Note that in his seminal paper \cite{Onsager1931} Onsager started with the specific case of thermoelectricity and generalized his concepts on coupled transports; However for a rigorous and complete description of thermoelectricity one should refer to Callen \cite{Callen1948} and Domenicali \cite{Domenicali1954}.

The article is organized as follows. In Section II we show that Onsager's reciprocal relations must be considered locally and that the extension of these relations to a global scale reveals the existence of an internal dissipative process. We discuss the consequences of such dissipations on finite-time thermodynamics and more particularly on efficiency at maximum power in Section III. Finally, in Section IV, basing our analysis on energy conservation, we discuss recent works which conclude to the possibility to overcome the traditional boundaries on efficiency imposed by finite-time thermodynamics in thermoelectric systems with broken time-reversal symmetry.

\section{\label{sec:localtoglobal} Local linearity and global dissipation}

\begin{figure}
	\centering
		\includegraphics[width=0.43\textwidth]{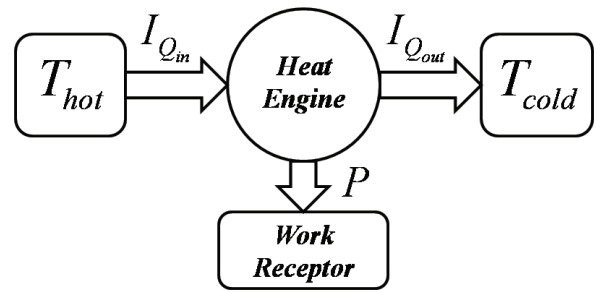}
	\caption{Thermodynamic picture of a heat engine.}
	\label{fig:figure1}
\end{figure}

Irreversible coupled thermodynamic processes may be described by a force-flux formalism \cite{Onsager1931}. In this framework, the first-order kinetic coefficients relating the fluxes to the affinities are supposed constant thus permitting a linear description of the system's thermodynamics. However, this linearity truly holds only locally. The hypothesis of local equilibrium is indeed a prerequisite to the definition of thermodynamic variables, which essentially describe equilibrium states; for a comprehensive discussion on this point, see Pottier's book \cite{Pottier2007}.

In the case of thermoelectricity, for which the coupled processes are the electrical charge transport and the heat transport, one may express Eq.~(\ref{Onsager}) on a local scale using the charge carrier current density $\vec{J}_N$ and the thermal flux density $\vec{J}_Q$ \cite{Callen1948}. The associated thermodynamic forces are then $X_1 = - \vec{\nabla} \mu / T$ and $X_2 = - \vec{\nabla} T / T^2$ with $\mu$ the local electrochemical potential and $T$ the local temperature. Eq.~(\ref{Onsager}) becomes:
\begin{equation}\label{OnsagerTE}
\left(
\begin{array}{c}
\vec{J}_N\\
\vec{J}_Q\\
\end{array}
\right)
=
\left(
\begin{array}{cc}
L_{11} ~ & L_{12} \\
L_{21}~ & ~L_{22} \\
\end{array}
\right)
\left(
\begin{array}{c}
- \vec{\nabla} \mu / T\\
- \vec{\nabla} T / T^2\\
\end{array}
\right).
\end{equation}

\noindent In his paper \cite{Callen1948}, Callen expressed the previous equation in terms of three appropriate quantities for a phenomenological description of thermoelectricity:
\begin{itemize}
  \item the electrical conductivity under isothermal condition $\sigma$, used in the local formulation of Ohm's law: $\vec{J} = \sigma \vec{E}$, where $\vec{E} = - \vec{\nabla} \mu / e$ is the local electrical field and $e$ is the elementary electric charge,

  \item the thermal conductivity under open-circuit condition($\equiv \vec{J} = \vec{0}$) $\kappa$, in Fourier's law: $\vec{J}_Q = \kappa \vec{\nabla} T$,
  
  \item the Seebeck coefficient $\alpha$, which characterizes the coupling between the electrical current density $\vec{J} = e \vec{J}_N$ and the thermal flux density $\vec{J}_Q$.
\end{itemize} 
\noindent The affinities are also modified to derive a form close to both Ohm's and Fourier's laws so that the force-flux relations for thermoelectricity are given by:
\begin{equation}\label{frcflxlocal}
\left(
\begin{array}{c}
\vec{J}\\
\vec{J}_Q\\
\end{array}
\right)
=
\left(
\begin{array}{cc}
\sigma ~ & ~\alpha \sigma \\
\alpha \sigma T~ & ~\alpha^2 \sigma T + \kappa \\
\end{array}
\right)
\left(
\begin{array}{c}
\vec{E}\\
- \vec{\nabla} T\\
\end{array}
\right)
\end{equation}
\noindent The relationships between the kinetic coefficients $L_{ij}$ and the traditional thermoelectric parameters $\sigma$, $\alpha$ and $\kappa$ are given in Ref.~\cite{Callen1948}. These force-flux relations are, as already stressed, linear only on a local scale: the coefficients in the matrix of thermoelectric coefficients (sometimes referred to as the conductivity matrix \cite{Goupil2011}) are given as an explicite function of the local temperature $T$ as $\sigma$, $\alpha$ and $\kappa$ are supposed constant.

The thermoelectric generator we consider is placed between two heat reservoirs, at temperatures $T_{\rm hot}$ and $T_{\rm cold}$ (with $T_{\rm hot} > T_{\rm cold}$), and delivers work to a load as depicted in Fig. \ref{fig:figure1}. To keep our calculations and analysis simple, we consider a one-dimensional structure associated with the spatial variable $x$. The hot reservoir is placed at $x = 0$ and the cold reservoir is placed at $x = l$. The section $A$ of the generator and the parameters $\sigma$, $\kappa$ and $\alpha$ are supposed constant. One last (trivial) assumption  that we make but which is often not explicit: the generator can exchange heat only with the reservoirs at their junctions, hence there are no heat leaks on the generator's sides.  

\subsection{From local to global scale}
To move from the local description of the intrinsic properties of the engine to a global description at the system level, we apply the principle of conservation of energy. The thermal current density is the sum of a convective term \cite{ApertetJPC} and a conductive term, respectively associated with the global movement of charge carriers that transports heat and Fourier's law:

\begin{equation}\label{convcondlocal}
\vec{J}_Q(x) = \alpha T(x) \vec{J} - \kappa \vec{\nabla} T (x)
\end{equation}

\noindent The local variation of the termal flux then reads:
\begin{equation}\label{heat2}
\vec{\nabla} \cdot \vec{J}_Q =  \vec{\nabla} T \cdot \alpha \vec{J} - \kappa \vec{\nabla}\cdot\vec{\nabla} T
\end{equation}

\noindent Now, since 
 
\begin{equation}\label{heat1}
\vec{\nabla} \cdot \vec{J}_Q = \vec{J} \cdot \vec{E}
\end{equation}

\noindent for a thermoelectric heat engine, a combination of Eqs.~(\ref{heat2}) and (\ref{heat1}), yields the the heat equation (sometimes called Domenicali's equation \cite{Domenicali1954b}) which reflects the conservation of energy:

\begin{equation}\label{heatequation}
\kappa \frac{{\rm d}^2T}{{\rm d}x^2} = - \frac{\vec{J}^2}{\sigma}
\end{equation}

From this stage, the boundary conditions must be taken into account: by placing the hot and cold heat reservoirs at $x=0$ and $x=l$ respectively, we set $T(0) = T_{\rm hot}$ and $T(l) = T_{\rm cold}$. The solution of Eq.~(\ref{heatequation}) gives the temperature gradient inside the structure:

\begin{equation}
\frac{{\rm d}T (x)}{{\rm d}x} = - \frac{T_{\rm hot} - T_{\rm cold}}{l} + \frac{\vec{J}^2 (l - 2x)}{2 \sigma \kappa}
\end{equation}

\noindent and injection of the above expression into Eq.~(\ref{convcondlocal}) yields the complete profile of the thermal current density along the thermoelectric generator. The incoming and outgoing thermal fluxes are defined as $I_{Q_{\rm in}}=A J_Q(0)$ and $I_{Q_{\rm out}}=A J_Q(l)$ respectively, so that they read:

\begin{eqnarray}\label{inandoutsym}
\nonumber
I_{Q_{\rm in}} &=& \alpha T_{\rm hot} I + K (T_{\rm hot} - T_{\rm cold}) - \frac{1}{2}R I^2 \\
&&~\\
\nonumber
I_{Q_{\rm out}} &=& \alpha T_{\rm cold} I + K (T_{\rm hot} - T_{\rm cold}) + \frac{1}{2}R I^2
\end{eqnarray}

\noindent with $K = \kappa  A /l$ being the thermal conductance, $R = l / (A \sigma)$ being the electrical resistance ,and $I = AJ$ being the electrical current .

\subsection{Discussion}
It is of importance to note that our reasoning made thus far would have been quite impractical had we considered Eq.~(\ref{Onsager}) only as the starting point. In Callen's study of thermoelectricity~\cite{Callen1948} the transport coefficients $\sigma$, $\kappa$, and $\alpha$ are constant and do not depend on the temperature while it may not be the case for the kinetic coefficients $L_{ij}$. As the output energy is directly related to the decrease of temperature experienced by the electrical charge carriers along the device, the knowledge of the explicit temperature dependence of each variable is essential to clearly express the condition of energy conservation. 

The power produced by the system is given by the difference between $I_{Q_{\rm in}}$ and $I_{Q_{\rm out}}$ :
\begin{equation}\label{power}
P = \alpha (T_{\rm hot} - T_{\rm cold}) I - R I^2
\end{equation}

\noindent The first term on the r.h.s of this expression reflects the fact that, as stated above, the energy extracted from the hot reservoir is linked to the movement of charge carriers from the hot to the cold reservoir; and an entropy per particle $\alpha$ is associated with each carrier. The second term, however, lowers the electrical power: part of the expected output power $\alpha (T_{\rm hot} - T_{\rm cold}) I$ is converted back to heat. This is the Joule heating, redistributed equally betwen the two reservoirs [see Eq.~(\ref{inandoutsym})].

Equations~(\ref{inandoutsym}) and (\ref{power}) exhibit terms that are quadratic in the electrical current $I$, and one issue that naturally comes to mind is the \emph{linear} character of the model. In a recent article, Izumida and Okuda presented their so-called minimally nonlinear model for heat engines adding quadratic terms in the linear expression of incoming and outgoing thermal fluxes~\cite{Izumida2012}. More precisely, they introduced a third flux in the Onsager formalism but actually this additional flux is merely a way to characterize the spatial variation of the thermal flux. We had already (though briefly) discussed the question of linearity in Ref.~\cite{Apertet2012a} but, here, we have clearly demonstrated that a thermodynamic analysis based a local linear model \emph{necessarily} yields a quadratic term associated with Joule heating in order to satisfy the principle of energy conservation. Once again we refer to Callen and Welton who associated a quadratic dissipation to linear systems~ \cite{Callen1951}:\\ \emph{The system may be said to be linear if the power dissipation is quadratic in the magnitude of the perturbation}.\\
\noindent Therefore, we argue that \emph{ad hoc} additions of nonlinear terms to models such as those usually discussed in the frame on linear irreversible thermodynamics to account for dissipation are misleading since these quadratic terms naturally appear in the equations as long as they are related to dissipation and ensure that the principle of conservation of energy in the conversion process is fully satisfied. An interesting discussion on the influence of friction, i.e., the mechanical analog of the electrical resistance, on the heat engine performance can be found in Ref.~\cite{Bizarro2012}. We also stress that for the particular case of thermoelectricity, the traditional model completely integrates the Joule heating in both the determination of the power and of the efficiency \cite{Ioffe1957}.

Finally, we analyze the confusion between continuous and discrete approaches to thermodynamic systems, which may be found in several recent papers: the thermodynamic affinities are often expressed as some spurious combination of the definitions for continuous media, i.e., gradients of thermodynamic potentials, and the definitions for the discrete description, i.e., differences of  thermodynamic potentials taken on the hot and cold sides of the engine. For clarity, we assign the following labels to the thermodynamic affinities $X$: discrete, local, and mixed; the latter expresses the combination of continuous and discrete descriptions. In the case of a discrete description of a thermoelectric engine, the affinities read $X_1^{(\rm discrete)} = - (\mu_{\rm hot} / T_{\rm hot}  - \mu_{\rm cold} / T_{\rm cold})$ and $X_2^{(\rm discrete)} = 1 / T_{\rm hot}  - 1 / T_{\rm cold}$ ($\mu_{\rm hot}$ and $\mu_{\rm cold}$ being respectively the electrochemical potential on the hot and cold ends of the engine) \cite{Callen1985}; in several works they are rather defined as $X_1^{(\rm mixed)} = - (\mu_{\rm hot}  - \mu_{\rm cold}) / T$ and $X_2^{(\rm mixed)} = - (T_{\rm hot}  - T_{\rm cold}) / T^2 $ where $T$ is the temperature of the system. These definitions should be compared to those at the local level, $X_1^{(\rm local)} = - \nabla \mu / T$ and $X_2^{(\rm local)} = - \nabla T / T^2$ \cite{Goupil2011}. We thus see that $X_1^{(\rm mixed)}$ and $X_2^{(\rm mixed)}$ are obtained by replacing the gradients of the local electrochemical potential and local temperature by the differences of these quantities taken on the hot and cold sides as though a discrete description was given. The major problem is that the temperature $T$ may no longer be considered as a local temperature and hence cannot be defined since the system is in an out-of-equilibrium state. In our view, this confusion is possibly one of the main reasons why  the effects of internal dissipations are neglected: this mixed, hence inappropriate, formulation of Onsager's relations leads to consider all systems as discrete and inasmuch as the distributions of the thermodynamic potentials within the system remains unknown, internal dissipations are completely neglected.

\section{Consequences on Finite Time Thermodynamics}
\subsection{Preliminary comments}
We just saw that internal dissipations derive from the linearity of the relationship between forces and fluxes at the local level. Now we seek what process ensures causality for this system. Since entropy production arises from internal dissipation, e.g., the Joule heating for thermoelectricity, the presence of finite thermal contact conductances in finite-time thermodynamics (FTT) models of non-ideal heat engines may no longer be viewed as mandatory: causality is ensured by the internal dissipation processes.

In this section, we propose an in-depth analysis of the recent literature on FTT and particularly the derivation of the efficiency at maximum power (EMP) in the light of our ``\emph{internal dissipation viewpoint}''. We assume that the engine works in the strong coupling regime: the coupling parameter $q = L_{12} / \sqrt{L_{11}L_{22}}$ equals $\pm1$ so that the two fluxes $J_1$ and $J_2$ are proportional \cite{Kedem1965, VandenBroeck2005}. The strong coupling assumption amounts to neglecting the \emph{bypass} heat, namely, the heat flowing from the hot to the cold reservoir without contributing to the energy conversion process (e.g., conduction through the walls of an engine's pipe). In thermoelectricity, in the strong coupling regime, thermal conduction contributes very little to energy conversion: $K \Delta T \ll \alpha \Delta T I$. As discussed in Ref.~\cite{Apertet2012c}, the bypass heat cannot be used to ensure causality: it is merely a form of  energy degradation that one must minimize.

\subsection{Endoreversibility vs. exoreversibility}

Most of the recent papers on EMP consider systems that are not endoreversible, but rather exoreversible: the thermal coupling between the engine and the heat reservoirs is assumed to be perfect (see, e.g., Refs. \cite{VandenBroeck2005, Esposito2010, Esposito2010b, Schmiedl2008, Apertet2012a, WangHe2012}). It is interesting to note that the Curzon and Ahlborn (CA) efficiency often is considered as a touchstone even if the endoreversible hypothesis used to derive it is no longer true. Yet, in 2008, Schmiedl and Seifert (SS) obtained a different expression for the EMP of a stochastic heat engine \cite{Schmiedl2008}:

\begin{equation}
\eta_{\rm SS} = \frac{\eta_{\rm C}}{2 - \gamma \eta_{\rm C}}
\end{equation} 

\noindent where $\gamma$ is a function of the entropy production repartition along the thermodynamic cycle with $0 < \gamma < 1$.

In a recent publication, we provided an explanation for the discrepancy between the two different expressions for the EMP that are $\eta_{\rm CA}$ and $\eta_{\rm SS}$ \cite{Apertet2012c}: studying the specific example of a thermoelectric generator, we demonstrated that $\eta_{\rm CA}$ is associated to the endoreversible configuration only while $\eta_{\rm SS}$ is associated to the exoreversible configuration only. Such a distinction between the system configurations, despite its primary importance from a theoretical point of view, seems to have been overlooked in many recent publications, even leading to the association of the CA efficiency with an exoreversible engine \cite{VandenBroeck2005}: Van den Broeck found that if each engine in the chain works at $\eta_{\rm C}/2$ (the \emph{linear} CA efficiency), then the whole system works at maximum power and has a global efficiency corresponding to the exact expression of CA efficiency: $\eta_{\rm CA} = 1 - \sqrt{T_{\rm cold} / T_{\rm hot}}$. Yet, we demonstrated that the maximum global power production is not always reached when all components work at maximum power \cite{Apertet2012a} and the expression derived \cite{VandenBroeck2005} does not necessarily correspond to the EMP. Furthermore Van den Broeck's result is intriguing since, as the exact expression of $\eta_{\rm CA}$ reproduces itself upon concatenation, it is surprising to obtain this exact expression for the global system without using the exact expression for each component; we thus believe that Van den Broeck's analysis is incomplete since a consideration for the internal dissipations is missing. The resulting confusion between endoreversible and exoreversible engines not only leads to the wrong expression of the EMP, i.e., $\eta_{\rm CA}$ instead of $\eta_{\rm SS}$, but also to some imprecisions in the interpretation of the thermodynamic cycle of the non-ideal Carnot engine as discussed below.

\subsection{Isothermal vs. adiabatic steps}

Carnot's cycle serves as a reference for the quantitative analysis of the impact of irreversibilities on a heat engine's operation. It is composed of two isothermal steps and two adiabatic steps.
In their paper Curzon and Ahlborn introduced entropy production during the isothermal steps only considering finite thermal resistances between the engine and the thermal baths while the adiabatic steps were assumed to be isentropic \cite{Curzon1975}. The recent studies based on Curzon and Ahlborn's analysis also assumed that the adiabatic steps are isentropic. However this is not mandatory since, as discussed in Ref.~\cite{Apertet2012c}, the introduction of dissipative thermal contacts is not the only way to ensure causality. Indeed, internal dissipation may also play this role. If only internal dissipations are accounted for, as for the case of exoreversible engines, entropy production may occur during the whole cycle and the adiabatic steps may no longer be considered as isentropic; hence the duration of these adiabatic steps should not be neglected.

Why is there entropy production during the whole cycle for exoreversible engines? The answer lies in the foundation of the finite-time thermodynamics: causality prevents a potential discontinuity between the heat engine and the load (e.g., continuity of the pressure on each side of the piston in the case of the classical Carnot engine described in Ref.~\cite{Carnot1824} or the continuity of the electrical potential in the case of thermoelectric heat engine). In all cases friction (in a general acception) must be present at all times to ensure the continuity of the potentials, hence the entropy production during the whole thermodynamic cycle over which a heat engine operates. Yet, if the external load's potential evolves infinitely slowly, the system may react as to always equilibrate with the external potential and no entropy may be produced if the friction depends on the actual state of the heat engine, as is the case for, e.g., viscous friction. However, note that this is not true in cases with constant friction forces such as that considered in Ref.~\cite{Bizarro2012AJP}: entropy production cannot vanish even in the limit of quasi-static processes and hence Carnot's efficiency cannot be reached.

In Ref.~\cite{Esposito2010} Esposito and coworkers derived a general expression of the EMP based on the so-called low dissipation assumption: the entropy production (the causes of which were not specified) in each isothermal step of a non-ideal Carnot cycle is inversely proportional to the duration of this step. Further, by considering that the total duration of the cycle is given by the sum of the durations of both isothermal steps, these authors neglected the contributions of the adiabatic steps to the entropy production however long the durations of these steps are. This latter consideration permits the assignment of an instantaneous character to the adiabatic steps, which corresponds to the most favorable configuration for power production. 

We distinguish endoreversibility from exoreversibility. In the former case neglecting adiabatic steps is justified since dissipations originate in heat exchanges with the thermal reservoirs rather than in the internal conversion processes. However, in the light of the discussion above, our view is that for the latter case, for which dissipations are mandatory during the whole cycle, the adiabatic steps should not be neglected, which implies that Esposito and coworkers' expression of the EMP is incomplete for endoreversible engines. 

The recent proposition of Wang and He to take into account the entropy productions occurring both during the isothermal and the adiabatic steps extends the study of Esposito and coworkers \cite{Esposito2010} to the exoreversible configuration \cite{WangHe2012}. Wang and He optimize four variables (while there were only two in Ref.~\cite{Esposito2010}) corresponding to the duration of each step of the thermodynamic cycle. The optimal value of each variable is given in \cite{WangHe2012} and we notice that the optimization of periodically driven exoreversible engines is actually quite tricky as the four steps must be considered. Yet, we demonstrate in the next section that the optimization of exoreversible autonomous engines remains simple. 

\subsection{Autonomous engine vs. periodically driven heat engines}

As emphasized in Ref.~\cite{Seifert2012}, classical engines are periodically driven: a time-dependent external control ensures that their operation follows their thermodynamic cycle, while the operation of autonomous engines corresponds to a non-equilibrium steady-state generated by externally-imposed time-independent boundary conditions (e.g., a load), which means that all steps occur virtually at the same time in the various parts of these engines. The main theoretical advantage of this kind of engine is that they can be described with Onsager's formalism \cite{Onsager1931}. To discuss the finite-time behavior of autonomous engines we consider the thermoelectric generator (other examples of both autonomous and periodically driven engines are considered in Ref.~\cite{Seifert2012}) and we give next a generalization from this particular case. Interestingly, the equivalent cycle experienced by electrical charge carriers is similar to a Carnot cycle: heat exchanges with thermal reservoirs indeed correspond to two isothermal steps while transport across both the thermoelectric generator and the electrical load may be viewed as two adiabatic steps.
 
As in Section \ref{sec:localtoglobal}, we obtain from the local Onsager relations and under the assumption of strong-coupling, the expressions for both the incoming and the outgoing thermal flux:
\begin{eqnarray}\label{toto}
\nonumber
I_{Q_{\rm in}} &=& \alpha T_{\rm hot} I  - \frac{1}{2}R I^2 \\
&&\\
\nonumber
I_{Q_{\rm out}} &=& \alpha T_{\rm cold} I  + \frac{1}{2}R I^2
\end{eqnarray}
\noindent The interest for these expressions is that they capture the whole behavior of the engine: they give the amount of heat exchanged with the thermal reservoirs during the equivalent isothermal steps of the cycle and contrary to Ref.~\cite{Esposito2010}, the contribution of the adiabatic steps is also accounted for through the Joule heating terms. As these latter result from the contribution of the whole device (as demonstrated in Section \ref{sec:localtoglobal}), we may now consider only the isothermal exchanges. In order to recover expressions similar to those of Ref.~\cite{Esposito2010} we consider the engine operation during a time $\tau$, so that $Q_{\rm in} = I_{Q_{\rm in}}\tau$ and $Q_{\rm out} = I_{Q_{\rm out}} \tau$. Since $I \tau = e N$  where $e$ is the elementary electric charge, $N$ is the number of particles entering the engine (from the load) during the hot isothermal step or (identically) leaving the engine (to the load) during the cold isothermal step, we get:
\begin{eqnarray}\label{heatinout}
\nonumber
Q_{\rm in} &=& T_{\rm hot}\left(\alpha e N - \frac{R e^2 N^2}{2 \tau T_{\rm hot}}\right) \\
&&\\
\nonumber
Q_{\rm out} &=& T_{\rm cold}\left(\alpha e N + \frac{R e^2 N^2}{2 \tau T_{\rm cold}}\right)
\end{eqnarray}

\noindent  One should then compare these expressions with those of Eq.~(5) of Ref.~\cite{Esposito2010}. Rewriting Eq.~(\ref{heatinout}) using the same notations yields:

\begin{eqnarray}\label{Espositolike}
\nonumber
Q_{\rm in} &=& T_{\rm hot} \left( \Delta S - \frac{\Sigma_{\rm hot}}{\tau}\right) \\
&&\\
\nonumber
Q_{\rm out} &=&  T_{\rm cold} \left( \Delta S + \frac{\Sigma_{\rm cold}}{\tau}\right)
\end{eqnarray}

\noindent and we see that the exchanged heat is the sum of a reversible contribution and an irreversible contribution for each isothermal step. The reversible entropy exchanged is $\Delta S = \alpha e N$, which is the product of the number of particles entering the heat engine during a time $\tau$ and $\alpha e$, sometimes referred to as the \emph{entropy per particle} \cite{Callen1948}. The irreversible contributions to the exchanged entropy are inversely proportional to $\tau$. The coefficients of proportionnality on the hot and cold sides are respectively $\Sigma_{\rm hot} =  e^2 N^2 R/2 T_{\rm hot}$ and $\Sigma_{\rm cold} = e^2 N^2 R/2 T_{\rm cold}$.

While the expressions derived here are very close to those of Esposito and coworkers ~\cite{Esposito2010}, the physical interpretation is completely different: we find that the produced entropy during the cycle is inversely proportionnal to the time $\tau$, thus recalling the low dissipation assumption, but this duration is not associated here to any specific thermodynamical process. Furthermore, the coefficients $\Sigma_{\rm hot}$ and $\Sigma_{\rm cold}$ do not represent the same physical reality as for Esposito and coworkers: they reflect only entropy production during the isothermal steps while in our case they account for the entropy produced during the whole cycle. It is interesting though that two different approaches on two different engine models (autonomous / periodically driven) yield the same mathematical expressions.

We now turn to the optimization of the autonomous engine. Assuming that the number of particles $N$ entering the engine during the time $\tau$ is constant, the variable used to perform the analysis of power maximization is $\tau$. We highlight the fact that one only needs a single variable to drive and hence to optimize an autonomous engine. This ascertainment is consistent with the fact that an autonomous engine depends only on a parameter imposed by the load. In the case of the thermoelectric generator this parameter is the value of the load resistance. Yet, controlling the load resistance $R_{\rm load}$ or $\tau$ is equivalent as they are related through the definition of the electrical current $I$ (that might also be used as the control parameter): as $I = eN/\tau = \alpha\Delta T /(R + R_{\rm load})$, we get $\tau = eN (R + R_{\rm load}) / (\alpha\Delta T)$.

The output power is defined as $P = (Q_{\rm in} -Q_{\rm out}) / \tau$, so that 

\begin{equation}
P = \frac{(T_{\rm hot}-T_{\rm cold}) \Delta S - (T_{\rm hot}\Sigma_{\rm hot} + T_{\rm cold}\Sigma_{\rm cold}) / \tau}{\tau}
\end{equation} 

\noindent  and the maximum power is then found for: 

\begin{equation}
\tau = 2 \frac{T_{\rm hot}\Sigma_{\rm hot} + T_{\rm cold}\Sigma_{\rm cold}}{(T_{\rm hot}-T_{\rm cold}) \Delta S}
\end{equation} 

\noindent Further, as $\eta = (Q_{\rm in} -Q_{\rm out}) / Q_{\rm in}$, we obtain the efficiency at maximum power:

\begin{equation}\label{etapmaxgenedemi}
\eta_{P_{\rm max}} = \frac{\eta_{\rm C}}{2 - T_{\rm hot}\Sigma_{\rm hot}\left(T_{\rm cold}\Sigma_{\rm cold} + T_{\rm hot}\Sigma_{\rm hot}\right)^{-1} \eta_{\rm C}}
\end{equation} 

\noindent which reduces to:
\begin{equation}
\eta_{P_{\rm max}} = \frac{\eta_C}{2 - \frac{1}{2}\eta_{\rm C}}
\end{equation} 
\noindent using the expressions for $\Sigma_{\rm hot}$ and $\Sigma_{\rm cold}$ obtained from Eqs. \eqref{heatinout} and \eqref{Espositolike}: $\Sigma_{\rm hot} =  e^2 N^2 R/2 T_{\rm hot}$ and $\Sigma_{\rm cold} = e^2 N^2 R/2 T_{\rm cold}$. The expression above is the Schmiedl-Seifert efficiency with $\gamma = 1/2$ (it was already derived for an exoreversible thermoelectric generator in Ref.~\cite{Apertet2012c}).

We now extend our analysis beyond the case of thermoelectric generators for which the heat internally produced by dissipation is equally distributed between the hot and cold reservoirs. We thus consider a general autonomous heat engine operating in the strong coupling regime. The proportion of internally produced heat flowing back to the hot thermal reservoirs is no longer set to $1/2$ but to a number $\beta$, the specific value of which is related to the constitutive equations governing the heat engine. Equation~(\ref{toto}) then becomes:

\begin{eqnarray}
\nonumber
I_{Q_{\rm in}} &=& \alpha T_{\rm hot} I  - \beta R I^2 \\
&&\\
\nonumber
I_{Q_{\rm out}} &=& \alpha T_{\rm cold} I  + (1-\beta) R I^2
\end{eqnarray}

\noindent Following exactly the same analysis as above, we obtain $Q_{\rm in}$ and $Q_{\rm out}$ as in Eq.~(\ref{Espositolike}) again, except that in the present case $\Sigma_{\rm hot}$ and $\Sigma_{\rm cold}$ read 

\begin{eqnarray}
\nonumber
\Sigma_{\rm hot} =  \frac{e^2 N^2 \beta R}{T_{\rm hot}}\\
&&\\
\nonumber
\Sigma_{\rm cold} = \frac{e^2 N^2 (1-\beta)R}{T_{\rm cold}}
\end{eqnarray}

\noindent Finally, inserting these two relations into Eq.~(\ref{etapmaxgenedemi}) yields the complete Schmiedl-Seifert efficiency:

\begin{equation}
\eta_{P_{\rm max}} = \frac{\eta_{\rm C}}{2 - \beta \eta_{\rm C}}
\end{equation} 

\noindent The major result associated with this derivation is the physical interpretation of the $\gamma$ factor in the Schmiedl-Seifert expression: as $\gamma = \beta$, it represents the proportion of heat produced by internal dissipation that is delivered to the hot thermal reservoir. It is consistent with the particular case of a thermoelectric generator for which half of the Joule heating is released in the hot reservoir. This result should be compared to the expression of $\gamma$ obtained by Schmiedl and Seifert (Eq.~(32) of Ref.~\cite{Schmiedl2008}): $\gamma$ was introduced using parameters called \emph{irreversible action} corresponding to $T_{\rm hot}\Sigma_{\rm hot}$ and $T_{\rm cold}\Sigma_{\rm cold}$. The expression they derived for $\gamma$ is $\gamma = 1 / \left[1 + \sqrt{\beta / (1-\beta)}\right]$ and thus does not correspond to our result as one might expect since the assumptions are not the same: in Ref.~\cite{Schmiedl2008} Schmiedl and Seifert used a periodically driven stochastic engine and they also neglected the contribution of the adiabatic steps just as Esposito and coworkers did \cite{CommentSS}.

To end this section we discuss the dependence of the Schmiedl-Seifert efficiency on the $\gamma$ factor (this discussion was initiated in Ref.~\cite{Apertet2012a}). For the exoreversible model, we may visualize the system as a classical energy converter, electrical or mechanical, for which Jacobi's theorem states that at maximum power the efficiency is half of the maximum value (here $\eta_{\rm C}/2$) \cite{VdB2012}. The power usually lost as heating may however be internally recycled as the considered engine converts heat into useful work. So, the incoming heat from the hot reservoir is converted once but as a part of the work produced is converted back to heat owing to the dissipations ensuring causality, the rejected heat may also pass through the conversion process once again if it is released on the hot side. The EMP is then slightly increased compared to $\eta_{\rm C}/2$, and the enhancement depends on the proportion of rejected heat allowed to undergo the conversion process again, i.e., it depends on $\gamma$. This situation may be viewed as an internal \emph{cogeneration} system where the dissipated heat may be converted in turn into useful power: the efficiency thus is slightly increased when heat is rejected to the hot reservoir.

\section{Energy conversion with broken time-reversal symmetry}
In this section, we discuss recent results concerning the possibility to obtain both a high energy conversion efficiency (close to $\eta_{\rm C}$), and a non-vanishing output power at the same time. This is quite intriguing as one of the main principle of FTT is the trade-off between efficiency and power. Recently, Benenti and coworkers claimed that a thermoelectric system operating under the condition of broken time-reversal symmetry may boast an EMP up to the Carnot efficiency \cite{Benenti2011}. To achieve this level of performance, the thermoelectric system must be submitted to a magnetic field $\vec{B}$ so that the time-reversal symmetry is broken and Onsager's relation $L_{12} = L_{21}$ is no longer true: in this case, it is the Onsager-Casimir relation $L_{12} ({\vec B}) = L_{21} (-\vec{B})$ that holds \cite{Callen1985}. If no magnetic field is applied, Onsager's relation is equivalent to the second Kelvin relation relating the Peltier coefficient $\Pi$ to the Seebeck coefficient $\alpha$: 
\begin{equation}
\Pi = \alpha T
\end{equation} 
\noindent with $T$ being the local temperature \cite{Thomson1882}. However, for a given applied magnetic field there is no reason, as stressed in Ref.~\cite{Benenti2011}, for this equality to hold.

To clearly present our calculations we express the Peltier coefficient as ${\Pi = \alpha_{\rm th} T}$, thus introducing a ``thermal'' thermopower, $\alpha_{\rm th}$. This form makes it clear that the transported heat per particle in the convection process, namely $\Pi$, depends on the local temperature. Consequently, to avoid confusion, we also define an ``electrical'' Seebeck coefficient: $\alpha_{\rm el}$. We thus split the thermoelectric response of the device into the temperature gradient represented by $\alpha_{\rm el}$ and the thermal response to the presence of an electrical current represented by the $\alpha_{\rm th}$. Hence, we reexpress Eq.~(\ref{frcflxlocal}) as:

\begin{equation}\label{frcflxassym}
\left(
\begin{array}{c}
\vec{J}\\
\vec{J}_Q\\
\end{array}
\right)
=
\left(
\begin{array}{cc}
\sigma ~ & ~\alpha_{\rm el} \sigma \\
\alpha_{\rm th} \sigma T~ & ~\alpha_{\rm el}\alpha_{\rm th} \sigma T + \kappa \\
\end{array}
\right)
\left(
\begin{array}{c}
\vec{E}\\
- \vec{\nabla} T\\
\end{array}
\right)
\end{equation}

\noindent now considering the possibility to introduce a dissymmetry as regards the electrical/thermal and the thermal/electrical couplings, i.e. ${\alpha_{\rm el} \neq \alpha_{\rm th}}$. Note that these may take negative or positive values.

\subsection{From local to global scale}
Just as we did in Section \ref{sec:localtoglobal}, we derive the expressions of the incoming and outgoing thermal fluxes based on the above linear relations and the appropriate boundary conditions which still are $T(0) = T_{\rm hot}$ and $T(l) = T_{\rm cold}$. The local thermal flux is given by:
\begin{equation}\label{convcondassym}
\vec{J}_Q(x) = \alpha_{\rm th} T(x) \vec{J} - \kappa \vec{\nabla} T (x)
\end{equation}

\noindent where the convective term now reflects the fact that each particle carries a quantity of heat $\Pi = \alpha_{\rm th} T$. The local energy balance is then:
\begin{equation}
\kappa \vec{\nabla}\cdot\vec{\nabla} T = - \frac{|\vec{J}|^2}{\sigma} +  (\alpha_{\rm th}-\alpha_{\rm el}) \vec{J}\cdot \vec{\nabla} T 
\end{equation}

\noindent We see the appearance of a second term proportional to the difference between the thermal and electrical Seebeck coefficients on the right-hand side. With this extra term, the temperature profile is no longer polynomial but depends on exponential functions of the spatial variable $x$:
\begin{eqnarray}
T(x) & = & T_{\rm hot} + \frac{|\vec{J}|x}{(\alpha_{\rm th}-\alpha_{\rm el})\sigma}+\frac{e^{\frac{(\alpha_{\rm th}-\alpha_{\rm el})|\vec{J}|}{\kappa}x}-1}{e^{\frac{(\alpha_{\rm th}-\alpha_{\rm el})|\vec{J}|}{\kappa}l}-1} \nonumber\\
& \times & \left[ - \Delta T - \frac{|\vec{J}|l}{(\alpha_{\rm th}-\alpha_{\rm el})\sigma}\right]
\end{eqnarray}

\noindent It is possible to obtain a much simpler form of the temperature profile by retaining the terms of its series expansion up to the second order in $x$. To do this, we use the fact that the current density is always smaller than $\alpha_{\rm el} \Delta T \sigma/ l$ in the generator regime. Because within the framework used in this article it is supposed that the temperature difference $\Delta T$ is very small, it follows that ${(\alpha_{\rm th}-\alpha_{\rm el}) |\vec{J}| l / \kappa \ll 1}$ and that the following result is justified: 
\begin{equation}
T(x) = T_{\rm hot} -  \frac{\Delta T}{l}x + \left[ \frac{|\vec{J}|^2}{2\sigma\kappa} + \frac{\Delta T}{l} \frac{(\alpha_{\rm th}-\alpha_{\rm el})|\vec{J}|}{2\kappa} \right] (l-x)x
\end{equation}

\noindent The derivative of $T$ with respect to $x$: 
\begin{equation}\label{dTexp}
\frac{{\rm d}T(x)}{{\rm d}x} = - \frac{\Delta T}{l} + \left[ \frac{|\vec{J}|^2}{2\sigma\kappa} + \frac{\Delta T}{l} \frac{(\alpha_{\rm th}-\alpha_{\rm el})|\vec{J}|}{2\kappa} \right] (l-2x)
\end{equation}

\noindent combined with of Eq.~(\ref{convcondassym}) yields the incoming and outgoing thermal fluxes for the whole system:
\begin{eqnarray}\label{inandoutdissym}
\nonumber
I_{Q_{\rm in}} &=& \alpha_{\rm th} T_{\rm hot} I + K \Delta T - \frac{R I^2}{2} - \frac{(\alpha_{\rm th}-\alpha_{\rm el}) \Delta T I}{2} \\
&&\\
\nonumber
I_{Q_{\rm out}} &=& \alpha_{\rm th} T_{\rm cold} I + K \Delta T + \frac{R I^2}{2} + \frac{(\alpha_{\rm th}-\alpha_{\rm el}) \Delta T I}{2}
\end{eqnarray}

In the absence of an applied magnetic field Eq.~(\ref{inandoutsym}) is recovered and $\alpha_{\rm th}=\alpha_{\rm el}$; the additional terms on the r.h.s. of Eq.~(\ref{inandoutdissym}), which appear for $B\neq 0$, are related to the broken time-reversal symmetry and it is of interest to see that the amount of thermal power ${(\alpha_{\rm th}-\alpha_{\rm el}) \Delta T I}$ is shared in two equal parts: one is added to the outgoing thermal flux whereas the other one is substracted from the incoming thermal flux (similarly to the Joule heating). We assume in the following discussion that $\alpha_{\rm th}$ and $\alpha_{\rm el}$ have the same sign; if not, the temperature difference $\Delta T$ would indeed imply a thermal flux flowing from the cold to the hot thermal reservoir in the generator regime, which would be in contradiction with Le Chatelier-Brown's principle.

To understand the physical meaning of the additional contributions to the thermal powers in Eq.~\eqref{inandoutdissym}, we consider the two cases $|\alpha_{\rm th}| > |\alpha_{\rm el}|$ \cite{discussion} and $|\alpha_{\rm th}| < |\alpha_{\rm el}|$. The former case corresponds to a situation where each charge carrier transports more heat than it would in the absence of the magnetic field so that the system needs to evacuate this excess; and the heat generated this way is also equally shared between the hot and the cold reservoirs. The latter case implies that the charge carriers do not transport enough heat: this lack is compensated by a higher thermal incoming flux and a smaller thermal outgoing flux. Finally, we see that the equipartition of generated heat remains valid.  

\subsection{Analysis of entropy production}
It is essential to check whether the process associated with the broken time-reversal symmetry always respects the second law of thermodynamics. To do so, we express the rate of entropy production, which is given by the difference between the rate of entropy increase in the cold reservoir and the rate of entropy decrease in the hot reservoir:  

\begin{equation}\label{entropiecree}
\frac{{\rm d}{\Delta S}_{\rm c}}{{\rm d}t} = \frac{I_{Q_{\rm out}}}{T_{\rm cold}} - \frac{I_{Q_{\rm in}}}{T_{\rm hot}}
\end{equation}

\noindent and using Eq.~(\ref{inandoutdissym}), we get: 

\begin{eqnarray}\label{entropiecree2}
\frac{{\rm d}{\Delta S}_{\rm c}}{{\rm d}t} &=& K \Delta T\left(\frac{\displaystyle 1}{\displaystyle T_{\rm cold}} - \frac{\displaystyle 1}{\displaystyle T_{\rm hot}}\right) + \frac{1}{2}R I^2 \left(\frac{\displaystyle 1}{\displaystyle T_{\rm cold}} + \frac{\displaystyle 1}{\displaystyle T_{\rm hot}}\right) \nonumber \\
&+& \frac{\displaystyle 1}{\displaystyle 2}\left(\alpha_{\rm th}-\alpha_{\rm el}\right) I \left(\frac{\displaystyle T_{\rm hot}}{\displaystyle T_{\rm cold}} - \frac{\displaystyle T_{\rm cold}}{\displaystyle T_{\rm hot}}\right)
\end{eqnarray}

\noindent The transfer of heat from the hot to the cold reservoir, with the related energy conversion, is physically allowed only if the rate of entropy production is positive. As the first two terms on the right-hand side of Eq.~(\ref{entropiecree2}), associated with heat by-pass and Joule heating respectively, are always positive, only the last term may reflect a behavior ``challenging'' the second law of thermodynamics. As above we may distinguish two cases:

\begin{itemize}
  \item If $|{\alpha_{\rm th}| \geq |\alpha_{\rm el}|}$, entropy can only be produced \cite{discussion}: the additional process caused by broken time-reversal symmetry is associated with pure dissipation just as Joule heating is.

  \item If $|{\alpha_{\rm th}| < |\alpha_{\rm el}|}$, the last term may be negative, so entropy destuction must always be compensated by entropy production associated with the first two terms of Eq.~(\ref{entropiecree2}).
\end{itemize}

\noindent To confirm that this latter condition is fulfilled for all possible working conditions in the generator regime, we determine the minimum entropy production rate and verify that this minimum is positive. The electrical current corresponding to this minimum is: 
\begin{equation}\label{Ientropymin}
I_{\rm min} = -\frac{(\alpha_{\rm th}-\alpha_{\rm el}) \Delta T}{2R}
\end{equation}

\noindent In the generator regime, the current lies between the open-circuit value, $I_{\rm oc} = 0$, and the short-circuit value, $I_{\rm sc} = \alpha_{\rm el} \Delta T / R$. It follows that in this range, the minimum entropy production rate corresponds to the open-circuit condition when $|{\alpha_{\rm th}| \geq |\alpha_{\rm el}|}$ but depends on the thermoelectric parameters when $|{\alpha_{\rm th}| < |\alpha_{\rm el}|}$. We take the opportunity here to highlight that the condition for maximum efficiency does not always correspond to the minimum entropy production rate as discussed in Ref.~\cite{Salamon2001}: if the thermal conductivity does not vanish, the maximum efficiency is reached for non-vanishing electrical current.

Using Eqs.~(\ref{entropiecree2}) and (\ref{Ientropymin}), we easily obtain the minimum entropy production rate, and the condition ${({{\rm d}{\Delta S}_{\rm c}/{\rm d}t})_{\rm min} \geq 0 }$ yields:
\begin{equation}\label{limiteK}
K \geq \frac{(\alpha_{\rm th}-\alpha_{\rm el})^2}{4R} \frac{T_{\rm hot} + T_{\rm cold}}{2}
\end{equation}

\noindent which amounts to defining a lower bound for the thermal conductance. This constraint is coherent with the  limitation of the figure of merit in the paper by Benenti \emph{et al.} \cite{Benenti2011}.

As we consider the case $|{\alpha_{\rm th}| < |\alpha_{\rm el}|}$, there is an additional internal process converting heat into electrical energy as described above. However, the energy transported by convection is already involved in the basic thermoelectric process and is not available for further conversion. The conversion associated with the broken time-reversal symmetry is thus based on the two sources of heat left: the Joule heat, which may be seen as a by-product of the thermoelectric conversion, and the by-pass heat, $K \Delta T$. The process is thermodynamically possible as long as there is enough heat to be converted; as Joule heating vanishes when reaching the open-circuit condition, the amount of by-pass heat must remain sufficient, which is equivalent to Eq.~(\ref{limiteK}).

\subsection{Power versus efficiency}
\begin{figure}
	\centering
		\includegraphics[width=0.50\textwidth]{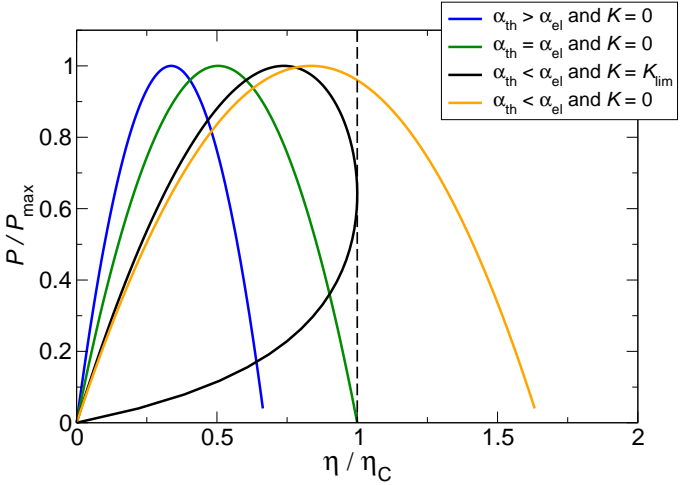}
	\caption{(Color online) Normalized power $P / P_{\rm max}$ vs. relative efficiency $\eta / \eta_{\rm C}$ under various working conditions. The curves are obtained by varying the electrical current $I$. The vertical dashed-dotted line represents the limit imposed by the second law, i.e. $\eta \leq \eta_{\rm C}$.}
	\label{fig:figure2}
\end{figure}

To complete our discussion on a thermoelectric generator operating under the condition of broken time-reversal symmetry, we also analyze the power versus efficiency curves. Using Eq.~(\ref{inandoutdissym}), we easily express the produced electrical power as a function of $I$: $P = I_{Q_{\rm in}} - I_{Q_{\rm out}} = \alpha_{\rm el} \Delta T I - R I^2$, and the efficiency: $\eta = P / I_{Q_{\rm in}}$. Figure~(\ref{fig:figure2}) displays the power $P$ as a function of the efficiency $\eta$ for various configurations. We first consider the strong coupling condition where the thermal conductivity vanishes, i.e., there is no thermal by-pass. For the situation without magnetic field ($|{\alpha_{\rm th}| = |\alpha_{\rm el}|}$), the maximum efficiency is, as expected, the Carnot efficiency, reached when the electrical current $I$ tends to zero. If $|{\alpha_{\rm th}| > |\alpha_{\rm el}|}$, the maximum efficiency is still reached for $I \rightarrow 0$, but the maximum value is lower than $\eta_{\rm C}$:

\begin{equation}\label{etamaxdissym}
\eta_{\rm max} = \frac{\eta_{\rm C}}{\frac{1}{2}\eta_{\rm C} + \left(1 - \frac{1}{2}\eta_{\rm C}\right)\frac{\displaystyle \alpha_{\rm th}}{\displaystyle \alpha_{\rm el}}}
\end{equation}

\noindent This difference arises from the fact that when approaching the open-circuit condition, the Joule heating process becomes negligible compared to the thermoelectric conversion, which is not the case for the process induced by broken time-reversal symmetry as this latter depends linearly on $I$ whereas Joule heating varies as $I^2$. If $|{\alpha_{\rm th}| < |\alpha_{\rm el}|}$, we see at first glance that the strong coupling assumption corresponds to a non-physical behavior since the generator may unrealistically boast efficiencies greater than the Carnot limit [Eq.~(\ref{etamaxdissym}) still holds].

To ensure that the second law of thermodynamics is not violated, the thermal conductance $K$ must satisfy Eq.~(\ref{limiteK}). Further if $K = K_{\rm lim} = (\alpha_{\rm th}-\alpha_{\rm el})^2 \overline{T} / 4R$, with $\overline{T}$ being the average temperature between the hot and the cold reservoirs, there exists a working condition for which the entropy production vanishes, so that the associated efficiency is $\eta_{\rm C}$. For this particular condition, the ``consumption of heat'' by the loss-of-time-symmetry induced process is exactly compensated by both the Joule heating and the thermal by-pass. The power vs. efficiency curve is then a closed loop, characteristic of non-vanishing thermal conductances, with a single point for which $\eta = \eta_{\rm C}$. The power associated to this point is different from zero, a fact that was surprising in the first place and gave hope for increasing thermoelectric conversion efficiency \cite{Benenti2011}. Note that the power further increases as the ratio ${|\alpha_{\rm el}| / |\alpha_{\rm th}|}$ increases.

\subsection{Efficiency at maximum power}
In addition to the maximum efficiency, we also derive the efficiency at maximum power, the pillar of the FTT analysis, for the thermogenerator operating under broken time-reversal symmetry. Since the maximum power is obtained for $I = \alpha_{\rm el} \Delta T / 2R$, we obtain the following general expression:
\begin{equation}\label{etaPmax}
\eta_{P_{\rm max}} = \frac{\eta_{\rm C}}{\left(2-\eta_{\rm C}\right)\frac{\displaystyle \alpha_{\rm th}}{\displaystyle \alpha_{\rm el}} +\frac{1}{2}\eta_{\rm C}+ \frac{\displaystyle 4 K R}{\displaystyle \alpha_{\rm el}^2 T_{\rm hot}}}
\end{equation}

\noindent When the thermal conductance is set to $K = K_{\rm lim} = (\alpha_{\rm th}-\alpha_{\rm el})^2 \overline{T} / 4R$, the previous expression is simplified:

\begin{equation}\label{etaPmaxKlim}
\eta_{P_{\rm max}} =  \frac{\eta_{\rm C}}{1 + \left( 1 - \frac{1}{2}\eta_{\rm C}\right) \left(\frac{\displaystyle \alpha_{\rm th}}{\displaystyle \alpha_{\rm el}}\right)^2 }
\end{equation}

\noindent The EMP $\eta_{P_{\rm max}}$ has a Lorentzian shape. Interestingly we recover the Schmiedl-Seifert expression for $\eta_{P_{\rm max}}$ in the classical situation where $B=0$, i.e., $\alpha_{\rm th}=\alpha_{\rm el}$, instead of the approximated value $\eta_{\rm C}/2$ obtained by Benenti and coworkers \cite{Benenti2011}.

\subsection{Discussion on Kelvin's relation}

If one considers that the thermal conductivity is independent of the value of $\alpha_{\rm th}$ and $\alpha_{\rm el}$, then the second law of thermodynamics imposes that $|\alpha_{\rm th}| \geq |\alpha_{\rm el}|$ for any finite value of the applied magnetic field $\vec{B}$ and in particular that:
\begin{equation}
\begin{array}{l}
|\alpha_{\rm th} (\vec{B})| \geq |\alpha_{\rm el} (\vec{B})|\\
|\alpha_{\rm th} (- \vec{B})| \geq |\alpha_{\rm el} (- \vec{B})|
\end{array}
\end{equation}

\noindent Further, considering the Onsager-Casimir relation, $\alpha_{\rm th} (\vec{B}) = \alpha_{\rm el} (-\vec{B})$, we get:

\begin{equation}
\begin{array}{l}
|\alpha_{\rm th} (\vec{B})| \geq |\alpha_{\rm el} (\vec{B})|\\
|\alpha_{\rm el} (\vec{B})| \geq |\alpha_{\rm th} (\vec{B})|
\end{array}
\end{equation}

\noindent and, since $\alpha_{\rm th}$ and $\alpha_{\rm el}$ have the same sign, we conclude that the only possible solution is 
\begin{equation}
\alpha_{\rm th} (\vec{B}) = \alpha_{\rm el} (\vec{B})
\end{equation}

\noindent which means that Kelvin's second relation holds even under the condition of broken time-reversal symmetry, and consequently that the Seebeck coefficient is an even function of the magnetic field. Such a conclusion was already given in Ref.~\cite{Tykodi1967}, but followed from different arguments. It is interesting to note that every material actually exhibits a Seebeck coefficent which is even with respect to $\vec{B}$ \cite{Wolfe1963}. 

\subsection{Mesoscopic realization}
Recently, Saito \emph{et al.} \cite{Saito2011} and S\'anchez and Serra \cite{Sanchez2011} proposed simultaneously a design of a mesoscopic system, which authorizes obtainment of the performances theoretically predicted in Ref.~\cite{Benenti2011}. This system, in addition to the two heat reservoirs, has a third probe/reservoir, which mimics the inelastic scattering processes supposed to enable a dissymmetry between the Peltier and the Seebeck coefficients. 

More recently, two articles by Balanchadran and coworkers \cite{Balachandran2013} and Brandner and coworkers \cite{Brandner2013} discussed the impact of the presence of a third connected lead on the behavior of a mesoscopic thermoelectric system. Constraining the third lead connection to the system to ensure current continuity, Brandner \emph{et al.} obtained new bounds for the maximum efficiency and the efficiency at maximum power, both lower than those obtained for the simpler system involving only two reservoirs studied here \cite{Brandner2013}. Our view is that the additional constraint on the kinetic coefficients is due to the fact that the third terminal may be viewed as a by-pass to the ballistic channel whose thermal conductance depends on the thermoelectric system's parameters, hence on $\alpha_{\rm th}$ and $\alpha_{\rm el}$. Imposing no net exchange of particles between the system and the third lead yields the condition: 
\begin{equation}
K \geq \frac{(\alpha_{\rm th}-\alpha_{\rm el})^2}{R} \frac{T_{\rm hot} + T_{\rm cold}}{2}
\end{equation}
\noindent which is a way of expressing Eq.~(15) of \cite{Brandner2013}, and constitutes a more stringent restriction on $K$ than that imposed by the second law of thermodynamics [see Eq.~(\ref{limiteK})]. In practice it seems that the introduction of a mechanism allowing thermopower asymmetry leads to such an increase of the thermal conductivity $K$ for the whole system that the second law is always satisfied.

\section{Summary and Conclusion}
We have demonstrated that internal dissipations must be considered in a linear model to satisfy the principle of energy conservation and to keep the model rigorously coherent. We stress that these dissipations are essential in the framework of finite-time thermodynamics for the analysis of systems which are not endoreversible. We also discussed the possibility of an additional internal mechanism related to a broken time-reversal symmetry proposed recently by Benenti and coworkers~\cite{Benenti2011}. It is our hope that our approach based on the Onsager-Callen formalism for thermoelectricity will allow a sharper view on the physical phenomena, which underlie the general principle of energy conservation.

\begin{acknowledgments}
Y. A. acknowledges financial support from the Minist\`ere de l'Enseignement Sup\'erieur et de la Recherche. We also acknowledge partial support of the French Agence Nationale de la Recherche (ANR), through the program ``Investissements d'Avenir''(ANR-10-LABX-09-01), LabEx EMC$^3$.
\end{acknowledgments}


\begin{thebibliography}{10}
\bibitem{discussion} For $\alpha_{\rm el} < 0$, the electrical current $I$ is also negative and the interpretation of Eq.~(\ref{inandoutdissym}) remains the same as in the case $\alpha_{\rm el} > 0$. Moreover, if $\alpha_{\rm el} > \alpha_{\rm th}$, the last term in Eq.~(\ref{entropiecree2}) is positive.
\end{thebibliography}
\end{document}